%% file: p.tex
\documentclass[sigconf,screen]{acmart}

\AtBeginDocument{%
  }

\copyrightyear{2026}
\acmYear{2026}
\setcopyright{cc}
\setcctype{by-nc-nd}
\acmConference[CCS '26]{Proceedings of the 2026 ACM SIGSAC Conference on Computer and Communications Security}{November 15--19, 2026}{The Hague, Netherlands}
\acmBooktitle{Proceedings of the 2026 ACM SIGSAC Conference on Computer and Communications Security (CCS '26), November 15--19, 2026, The Hague, Netherlands}
\acmDOI{10.1145/3830454.3846577}
\acmISBN{979-8-4007-2871-6/2026/11}

\usepackage{url}

\usepackage{epsfig,endnotes}

\include{pkgs}
\usepackage{tcolorbox}

\newcommand{\sys}{\mbox{\textsc{Tirith}}\xspace}

\usepackage{hyperref}
\usepackage{adjustbox}
\usepackage{enumitem}
\usepackage{threeparttable}

\newcommand{\AD}[1]{\ifdefined\SHOWCOMMENTS\textcolor{red!70}{AD: #1}
\else{}\fi}
\newcommand{\CQ}[1]{\ifdefined\SHOWCOMMENTS\textcolor{blue!60}{CQ: #1}
\else{}\fi}

\newcommand{\SL}[1]{\ifdefined\SHOWCOMMENTS\textcolor{magenta!90}{SL: #1}
\else{}\fi}

\newcommand{\Revision}[2]{{\ifdefined\HIGHLIGHTREVISION{\color{blue}{\bf #1} {#2}}\else{{#2}}\fi}}

\input{cmds}

\begin{document}
\input{hdr}

\input{abstract}

\date{}


\maketitle
\sloppy 

\input{intro}
\input{motivation}
\input{background}

\input{design}
\input{case-study}
\input{security}
\input{implementation}
\input{performance}
\input{relwk}
\input{conclusion}

\bibliographystyle{plain}
\bibliography{p,sslab,conf}

\appendix

\section{Open Science}

All our artifacts are available at: 
\url{https://github.com/ASTERISC-Release/Tirith}.

    
    
    

%



\section{Generative AI Usage}
Large Language Models (LLMs) were used for copy-edit purposes to improve the clarity and grammar of our paper writing.
Additionally, we used LLMs to help debug our implementation.
All usage of LLMs was reviewed by the paper authors.

\end{document}

%% file: pkgs.tex
\usepackage{amsmath,amsopn,amsthm}
\usepackage{booktabs}
\usepackage{multirow}
\usepackage{subcaption}
\usepackage{endnotes}
\usepackage{microtype}
\usepackage{xspace}
\usepackage{graphicx}
\usepackage{fancyvrb}
\usepackage[most]{tcolorbox}
\usepackage{listings}
\usepackage{svg}
\tcbuselibrary{listings, skins} 
\usepackage{fp}
\usepackage[mode=text]{siunitx}
\usepackage[utf8]{inputenc}
\usetikzlibrary{calc}
\usepackage{booktabs}
\usepackage[utf8]{inputenc}

%% file: cmds.tex
\usepackage[T1]{fontenc}

\newcommand{\cc}[1]{\mbox{\texttt{#1}}}

\newcommand{\x}{\ding{55}\xspace}

\input{code/fmt}

\def\Snospace~{\S{}}

\renewcommand{\x}{$\times$\xspace}

\usepackage{courier}
\usepackage[all]{nowidow}

\newif\ifdraft\drafttrue
\newif\ifnotes\notestrue
\ifdraft\else\notesfalse\fi

\input{glyphtounicode}
\newcolumntype{R}[1]{>{\raggedleft\let\newline\\\arraybackslash\hspace{0pt}}p{#1}}

\newcommand{\includepdf}[1]{
  \includegraphics[width=\columnwidth]{#1}
}

\newcommand{\squishlist}{
\begin{itemize}[noitemsep,nolistsep]
  \setlength{\itemsep}{-0pt}
}
\newcommand{\squishend}{
  \end{itemize}
}

\usepackage{tikz}

\newcommand{\sysgame}{p\textsc{VM}\xspace}

\definecolor{colorextra}{RGB}{234,232,218}

\newtcolorbox{conclusionbox}{colback=gray!8,colframe=black,width=\linewidth,arc=0.6mm, boxrule=0.8pt, left=1mm,right=1mm,top=1mm,bottom=1mm}

\newcommand*\WC[1]{%
\begin{tikzpicture}[baseline=(C.base)]
\node[draw,circle,inner sep=0.2pt](C) {#1};
\end{tikzpicture}}

\newcommand*\BC[1]{%
\begin{tikzpicture}[baseline=(C.base)]
\node[draw,circle,fill=black,inner sep=0.2pt](C) {\textcolor{white}{#1}};
\end{tikzpicture}}

\usepackage{xstring}
\newcommand{\PP}[1]{
\vspace*{1.2pt}
\noindent{\bf \IfEndWith{#1}{.}{#1}{#1.}}
}

\newcommand{\PPn}[1]{
\vspace*{1.2pt}
\noindent{\bf #1}
}

\newcommand{\CC}[1]{
\vspace*{1.2pt}
\textit{\IfEndWith{#1}{.}{#1}{#1.}}
}

\usepackage{pifont}

\makeatletter
\patchcmd{\ttlh@hang}{\parindent\z@}{\parindent\z@\leavevmode}{}{}
\patchcmd{\ttlh@hang}{\noindent}{}{}{}
\makeatother

\newtcolorbox{mybox}
{
  before skip=2mm,
  boxsep=0.5mm,
  top=1mm,
  bottom=1mm,
}

\newenvironment{packeditemize}{
\begin{list}{$\bullet$}{
\setlength{\itemsep}{2pt}
\setlength{\labelwidth}{8pt}
\setlength{\leftmargin}{10pt}
\setlength{\labelsep}{3pt}
\setlength{\listparindent}{\parindent}
\setlength{\parsep}{1.5pt}
\setlength{\parskip}{1.5pt}
\setlength{\topsep}{1.5pt}}}{\end{list}}


%% file: hdr.tex

\title{You Shall Not Pass into Ring-0! A User Privacy-Friendly Anti-Cheat Architecture for Personal Computers}



\author{Santosh Gokul Narayanan}
\orcid{0009-0002-6608-3032}
\authornote{Both authors contributed equally to this work. Work completed while both authors were students at Arizona State University.}
\affiliation{  \institution{Nokia of America Corporation}
  \city{Sunnyvale}
  \country{USA}
}
\email{santosh.narayanan@nokia.com}

\author{Giovanni Paladino}
\orcid{0009-0005-7878-3097}
\authornotemark[1]
\affiliation{  \institution{New York University}
  \city{New York}
  \country{USA}
}
\email{gpp9131@nyu.edu}

\author{Chuqi Zhang}
\orcid{0009-0006-3550-696X}
\affiliation{  
\institution{National University of Singapore}
  \city{Singapore}
  \country{Singapore}
}
\email{chuqiz@u.nus.edu}

\author{Sangho Lee}
\orcid{0000-0002-0412-7768}
\affiliation{  \institution{Microsoft Research}
  \city{Redmond}
  \country{USA}
}
\email{sangho.lee@microsoft.com}

\author{Zhenkai Liang}
\orcid{0000-0001-7138-5030}
\affiliation{  \institution{National University of Singapore}
  \city{Singapore}
  \country{Singapore}
}
\email{liangzk@nus.edu.sg}

\author{Adil Ahmad}
\orcid{0009-0002-4097-3205}
\authornote{Corresponding author.}
\affiliation{  
  \institution{Arizona State University}
  \city{Tempe}
  \country{USA}
}
\email{adil.ahmad@asu.edu}

%% file: abstract.tex
\begin{abstract}

Kernel-level anti-cheats are effective against malicious player behavior in competitive video games, but raise significant user privacy concerns regarding installing unverifiable components at privileged modes~(i.e., ring-0 in x86).
While existing research has focused on improving the effectiveness of anti-cheats, the user privacy concern has been largely ignored.
\sys is an anti-cheat architecture that addresses this problem using two key ideas.
First, instead of running video games within regular processes that players (as root admins) have control over, \sys executes video games in Protected Virtual Machines that naturally sandbox computations from untrusted admins.
Second, to monitor user behavior outside the sandbox (e.g., see if they are running malicious drivers), \sys leverages a virtualization monitor that is trusted by both players and developers.
Together, these ideas remove the need to run untrusted kernel-level anti-cheats, while providing the same level of protection compared to such solutions against a wide-range of common cheating mechanisms.
The main challenge we face in implementing these ideas, however, is that the existing software stack for virtual machines is not designed to run video games 
and creates significant security and performance problems.
We address these problems by proposing a security-focused Library OS kernel for games and an efficient graphics sharing pipeline for near-native rendering and display performance.
In summary, without compromising on cheating behavior detection or performance, this work makes user privacy a first-class citizen in personal computers. 

\end{abstract}

%% file: intro.tex
\section{Introduction}
\label{s:intro}

With revenue exceeding \$195 billion in 2025~\cite{newzoo_2025_review}, online gaming is the largest entertainment sector in today's world.
This revenue is driven by titles such as Fortnite, Apex Legends, CS2, and Valorant where players compete with each other for prestige, prizes, and even online followers through streaming platforms.
%
%
Sadly, this aspect leads to cheating---professional teams, popular streamers, and casual players alike are routinely caught using cheats sold openly by vendors such as EngineOwning~\cite{engineowning_detection_2024}.
Such cheating behaviors disincentivize honest players, an aspect that has motivated game developers to invest heavily in countermeasures.
%

%
The prevalent countermeasure today
is to deploy a class of solutions colloquially referred to as {\em kernel-level anti-cheats} by players~(\autoref{s:kernel-level-anticheat}).
As the name implies, such solutions install a privileged kernel driver on a player's computer, allowing them to monitor the system environment and prevent potentially malicious behavior (e.g., installing unsigned kernel modules and blocking debuggers from attaching to the game process).
This driver typically operates in conjunction with a user-level component (running in the game process) to collect additional information and send it to the server for complete cheating behavior detection.
%
%

Unfortunately, kernel-level anti-cheats install untrusted (i.e., developer-controlled) software in privileged domain, leading to severe privacy risks for players: such anti-cheats have been found to collect telemetry far beyond gameplay, and in some cases ship spyware/rootkit-like behavior~\cite{Dorner2024examination}.
%
\Revision{(R6)}{They also complicate security aspects for game developers. 
As modern kernels like Linux continue to grow larger and more complex, vulnerabilities in the kernel's codebase allows dishonest players to circumvent anti-cheat mechanisms (e.g., by tampering with anti-cheat modules after kernel compromise through a vulnerable signed driver~\cite{bitdefender_byovd}).}
While there has been significant recent research on anti-cheat solutions~\cite{botscreen,blackmirror,avm,invisiblecloak}, it has focused on two orthogonal lines \Revision{(R6)}{that do not adequately address the dual privacy-security} risks introduced by kernel-level anti-cheats~(\autoref{s:existing-work}).
%
Specifically, the first line of research focuses on leveraging machine learning-based techniques to improve the effectiveness of detecting attacks like aimbots and wallhacks.
The second line of research focuses on only improving the security of anti-cheat detectors by executing them within trusted execution environments~(e.g., Intel SGX).
%

%


We present \sys\footnote{In J.R.R. Tolkien's Sindarin language, Tirith means ``guard,'' ``watch,'' or ``vigilance.''}, a privacy-friendly solution 
%
that obviates the need to run kernel-level anti-cheat components, 
\Revision{(R6)}{while also strengthening anti-cheat security against kernel compromises.}
%
%
The solution leverages two key ideas:

\begin{packeditemize}
    \item {\em Virtualization-Based Sandboxing}: Instead of running games in processes that users control, each game is isolated into a {\em Protected Virtual Machine} (PVM), a hardware-isolated guest environment provisioned by the OS vendor's first-party trusted hypervisor and increasingly available on consumer machines~\cite{vbs,pkvm}.
    %
    This provides natural two-way isolation of CPU and memory contexts between guest and host environments.

    \item {\em Split Responsibility Architecture:}
    A PVM must still share devices with the host. A player who is running a compromised platform (e.g., with unsigned drivers) can abuse this sharing to launch attacks against games. 
    %
    We split the responsibility of tracking whether the platform is trustworthy or not to the trusted and independent PVM hypervisor, instead of developer logic, and allow the developer to query this before starting game sessions.
    
    %

\end{packeditemize}

To understand the feasibility of our approach, we ran a case-study on running video games using the state-of-the-art virtualization stack available on Linux~(\autoref{s:virtualization}).
We found two main problems.
First, existing implementations all leverage a full-fledged OS like Linux within the virtualized environment, but this approach is both resource-intensive and raises security problems since Linux-based guest virtual machines require many interfaces to the outside world making it hard to reason about guest protection and host privacy.
Second, all existing paravirtualization techniques to share the GPU with guest virtual machines incur prohibitive overheads running video games, since they require expensive coordination and slow message-passing between the guest and host contexts.
%
%

%
To address the first problem, we propose a {\em minimized game-centric library OS kernel} that hosts only the game and its developer's anti-cheat module inside the PVM~(\autoref{s:design:libos}).
The kernel is extended with a virtual device subsystem exposing a shim DRM render node, an event device for input, a sound device, and a platform attestation device, so that existing games run unmodified.
Game binaries and assets are loaded from a developer-signed manifest whose root hash is part of the boot measurement, while less latency-critical channels (network, sound, storage) attach via standard {\em virtio} interfaces.
The LibOS also exposes hypercalls so the in-guest anti-cheat can collaboratively invoke trusted-monitor primitives, e.g., to attest the host's driver-admission policy or query host platform measurements.
We address the second problem with a {\em native shared GEM (Graphics Execution Manager) context}~(\autoref{s:design:rendering}) that avoids the graphics command stream encode/decode and VM-exit costs of conventional GPU paravirtualization.
The host VMM allocates per-process GEM buffers from the host DRM render node, and the trusted hypervisor projects them into a pre-reserved region of guest physical memory under a write-combining cache policy.
Inside the guest, the game and its OpenGL stack issue draw calls directly against these shared buffers with no frame copy or format conversion.
Only the control plane (\cc{ioctl} and graphics syscalls on the render node) is forwarded to the host to interact with the physical GPU, through an exception-less shared-memory channel.
The host then displays the frames (which already live in a host-mapped GEM buffer) at native cost using a bridged windowing approach.
%

%
%

%
We implemented a \sys prototype for a Linux-based gaming environment using KVM as the type-1 hypervisor backend.
The game-centric kernel is built on the Gramine Library OS~\cite{gramine-tdx, graphene} and extended with the virtual device subsystem, manifest-based asset integrity, exception-less host RPC channel, and trusted-monitor hypercalls (platform measurement, driver-policy attestation, attestation digests) described above.
The native shared-GEM rendering pipeline supports OpenGL 4.6 applications via the EGL specification, and the bridged display/input pipeline targets X11.
Our prototype is available at \url{https://github.com/ASTERISC-Release/Tirith}.
Using our prototype, we conduct a structured security analysis~(\autoref{s:security-analysis}) of \sys from both sides of the trust boundary.
We show that the player's host privacy is preserved by hypervisor-based sandboxing together with a small, mediated set of guest-host interfaces, and that game security is upheld with the same anti-cheat primitives kernel-level anti-cheats rely on today --- now enforced inside the guest where the developer maintains full control.
We further illustrate the analysis against three real-world cheat families (wallhacks, aimbots/triggerbots, and map hacks), each defeated by a combination of in-guest and monitor-delegated primitives.
We also provide a detailed performance evaluation of \sys using benchmarks and real-world video games~(\autoref{s:perf}).
%
Compared with the state-of-the-art paravirtualized stack (VirGL and DRM-Native),
\sys delivers near-native rendering throughput and input latency while reducing guest CPU and memory utilization, narrowing the gap to non-virtualized Linux to within a small constant factor across all three games.
%
   %
Concretely, our evaluation shows that \sys{} achieves average FPS and $1\%$ lows within $4.1$-$5.2\%$ and $16.5$-$18.2\%$ of native respectively, which is $2.4$-$2.8\times$ and $3$-$3.8\times$ better than existing paravirtualization solutions.
In conclusion, our work demonstrates that strong user privacy is achievable in modern computers, without compromising on cheat prevention or performance.

%% file: motivation.tex
\section{Motivation}
\label{s:motivation}


\subsection{Kernel-Level Anti-Cheats and Risks}
\label{s:kernel-level-anticheat}


Malicious players can attempt to interfere with game execution by modifying memory, injecting code, attaching debuggers, loading malicious drivers, or bypassing user-space enforcement mechanisms~\cite{collins2024anticheat}.
%
To address this problem, 
the current design of anti-cheats typically involves both a user-level component and a {\em privileged kernel component}.
Their primitives are shown in \autoref{t:primitive-vbs-repartition}.
The user-level component is game-specific and deployed alongside the game client. 
It is responsible for tasks like collecting information from the game and sending it to servers for analysis (e.g., using machine learning to detect aimbots~\cite{avm,botscreen,kalra2018blockchain-based}).
In contrast, the kernel component is installed as drivers or modules~\cite{riot_vanguard_2024,battleye_about,eac,faceit,avm}.
It offers {\em global system visibility and control}, allowing the anti-cheat to observe and regulate behaviors relevant to game integrity, including memory access, library injection, driver loading, debugging activity, and interactions with input and display paths.
%
%
\autoref{t:primitive-vbs-repartition} illustrates the  responsibilities of user and kernel anti-cheat components.

Famous examples of anti-cheat software that follow the aforementioned design include Riot Vanguard~\cite{riot_vanguard_2024}, Easy Anti-Cheat (EAC)~\cite{eac}, FACEIT~\cite{faceit}, and BattlEye~\cite{battleye_about}, colloquially referred to as {\em Kernel-Level Anti-Cheats} by users.
These systems are {highly-resistant to tampering}, since software running at high privilege can better monitor its own integrity and resist attempts to disable or bypass its enforcement logic.
For example, Vanguard integrates with secure boot and loads early during system startup to establish privileged protection and strengthen its integrity~\cite{riot_vanguard_2024}.
Sadly, {\bf kernel-level anti-cheats introduce severe privacy risks for \Revision{(R6)}{honest players, while also providing limited security against powerful dishonest players}}.
%
Players are asked to allow closed-source, developer-controlled components to execute with kernel privileges.
With such privileges, anti-cheat software may access host data unrelated to gameplay, interfere with benign applications, and affect normal system behavior~\cite{riot_vanguard_faq}.
For example, Vanguard interposes UI windows to detect display-based cheat overlays, but this capability has also raised concerns because it enables full-screen screenshot capture~\cite{all-your-pixel}.
%
%
Some widely deployed kernel-level anti-cheat systems have also been criticized as ``rootkit-like''~\cite{Dorner2024examination}.
Such systems may conceal their presence and resist inspection or removal, while retaining broad capabilities for remote access and data collection~\cite{red-shell}.
%
%
\Revision{(R6)}{On the security side, kernel anti-cheats can also be circumvented by abusing core kernel vulnerabilities~\cite{dirtycred,retspill} or device drivers~\cite{bitdefender_byovd} to achieve full kernel compromise. 
Consider the Genshin Impact attack, where a signed vulnerable driver was used by a ransomware actor to bypass privileges and kill antivirus processes and services during a mass-ransomware deployment~\cite{trendmicro_genshin_byovd_2022}.
This attack shows that vulnerable drivers remain a realistic threat vector even against privileged software. 
}

\input{tables/primitives-VBS}

\subsection{Scope of Existing Anti-Cheat Research}
\label{s:existing-work}



Prior work has focused on {\em improving the detection or prevention of cheating behavior}, but has not addressed the privacy risks introduced by kernel-level anti-cheat software itself.
%
%
These solutions can be broadly grouped into the categories below.

%

The first category improves {cheat detection or defense against specific cheat pipelines}.
These approaches collect richer evidence or directly target particular cheating behaviors.
For example, AVM~\cite{avm} makes execution tamper-evident and replayable, enabling suspicious game behavior to be audited against recorded execution history.
Other systems use behavioral signals and machine learning to detect cheating patterns such as aimbots~\cite{alayed-et-al, bauman-et-al, vacnet}.
Invisibility Cloak~\cite{invisiblecloak} proactively perturbs rendered frames to make visual aimbots unreliable while preserving the normal visual experience for human players.
Although these techniques improve cheat detection or defend against specific attacks, they do not address whether proprietary anti-cheat should hold privileged access on the host.


The second category uses trusted execution environments (TEEs) to enable lightweight countermeasures against particular cheating behaviors while protecting the defense logic itself from an untrusted host.
For example, BlackMirror~\cite{blackmirror} uses Intel SGX to protect hidden game state and performs trusted visibility testing before rendering, thereby preventing wallhacks.
BotScreen~\cite{botscreen} places a pre-trained model inside an SGX enclave to detect aimbots on the client side.
They show that trusted execution can harden selected portions of the game protection pipeline.
However, they remain {\em protection-centric point solutions}: each targets a specific cheat surface, and user-space TEEs such as SGX do not replace the broader range of privileged enforcement mechanisms still relied upon by today’s anti-cheat systems in practice.
TZMon~\cite{jeon2021tzmon} relocates mobile game anti-cheat into the ARM TrustZone secure-world kernel, hardening it against a compromised Android OS.
The anti-cheat itself, however, still holds privileges over the normal-world OS to inspect game state, retaining the same trust issue.

\section{Goal and Approach}
\label{s:overview}

\begin{figure}[t]
    \footnotesize
    \begin{center}
      \centering
    \includegraphics[width=\columnwidth]{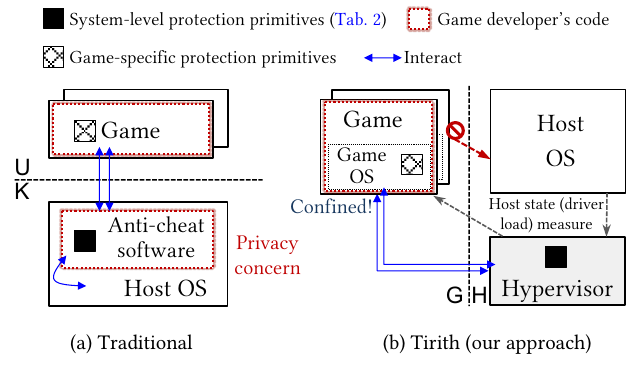}
      \caption {Overview of the VM-based sandboxing with split responsibility architecture (G: guest; H: host).\AD{looks like you are loading a driver in the hypervisor.}}
     \label{f:insight}
    \end{center}
\end{figure}

Our goal is to design a privacy-friendly anti-cheat architecture for Linux-based computers that does not rely on developer-controlled privileged components.
%
%
%
We target Linux because kernel-level anti-cheats are notably absent, even as gaming on the platform continues to grow in popularity.
Anti-cheat developers themselves argue this gap is not incidental: the Linux kernel restricts proprietary modules from tightly coupling with the kernel through GPL-only symbols, and any user can compile a custom kernel that hides cheats below an anti-cheat's view~\cite{conway2025protonlinux,fedora_anticheat_discussion}.
This makes Linux precisely the right platform on which to rethink anti-cheat architecture from the ground up and treat host privacy as a first-class citizen.
Nonetheless, the resulting design also ports to other platforms like Windows.
Our approach (\autoref{s:approach:sandbox} - \autoref{s:approach:split}) is shown in \autoref{f:insight}.

\subsection{Virtualization-Based Sandboxing}
\label{s:approach:sandbox}

Games today are executed in regular user-level processes, even though their threat model is distinct from typical applications that execute in processes.
For typical applications, the user (who has {\em root}/{\em admin} privileges on the computer) is assumed to be trusted.
This assumption simply does not hold true for competitive games where the user is assumed to be actively trying to tamper with the game.
Therefore, our first insight is that {\em the process abstraction is fundamentally unsuited to running competitive games}.

Based on the aforementioned insight, we propose to leverage an emerging abstraction in personal computers that naturally provides protection to computations against a root user, namely the {\em Protected Virtual Machine} (PVM).
This abstraction is realized using a type-1 hypervisor that executes on the computer and isolates a guest virtual machine's context from an untrusted host.
Modern commodity platforms increasingly implement a PVM abstraction layer, including Protected KVM on Linux/Android, and Virtualization-Based Security~(VBS) Enclave on Windows~\cite{vbs,pkvm,pkvm-android}.
%


The PVM abstraction provides several desirable properties required for our architecture.
(i)~{\em First-party and signed}: the hypervisor is provisioned by the OS vendor and launched as part of secure boot, requiring no trust in any third-party developer code.
(ii)~{\em Higher privileged than the kernel}: the hypervisor cannot be disabled or bypassed by software running on top, including the player acting as root.
(iii)~{\em Strong two-way guest isolation}: it hosts PVMs whose memory, CPU state, and devices are hardware-isolated from the host, so that neither the host nor the guest can inspect or tamper with the other.
(iv)~{\em Remote attestation}: Both the hypervisor and protected guest can be measured to produce attestation reports that a remote party (e.g., a game server) can verify.

%
Based on these properties, each game can be executed inside its own PVM guest to provide strong isolation from the host.
In the guest, the game and the developer's anti-cheat logic run with kernel level privilege over the game environment, giving the developer full control over the guest for monitoring and enforcement.
The same isolation works in the opposite direction.
The developer's code is confined to PVM: it cannot read host memory, spy on unrelated applications or data, capture arbitrary screen contents, or persist on the machine after a session ends.\AD{pass}

\subsection{Split Responsibility Architecture}
\label{s:approach:split}

While Protected Virtual Machines (PVMs) provide isolation and sandboxing to computations running on the CPU, they cannot {\em alone} fully-protect computations like video games that leverage external devices like GPUs and input devices (e.g., keyboard).
In particular, external devices are shared with the host using device-sharing primitives~(discussed in~\autoref{s:virtualization}), allowing an untrusted host to launch attacks.
%
For example, a root user can install malicious drivers that manipulates the framebuffer for display, or tamper with game binaries/assets on the host-backed guest filesystem to substitute modified resources.
Therefore, {\em complete video game protection requires some degree of visibility and control over the host}.

%



Fortunately, we observe that the required host visibility and control can be achieved using the trusted PVM hypervisor~\cite{secvisor}.
%
%
Taking Windows VBS as an example, the hypervisor implements secure boot and measured launch establish a trusted boot chain. 
Moreover, the hypervisor governs driver installation and decides which kernel modules can load, as well as leverages the IOMMU and interrupt remapping mediate what devices can touch protected memory.
It also supports remote attestation to expose the resulting state to a verifier.
In fact, we are already seeing the hypervisor becoming an important cornerstone of anti-cheat protection. For instance, the developers of Call of Duty: Black Ops 7 already incorporate VBS driver verification into their anti-cheat pipeline to strengthen protection against host-resident cheats~\cite{cod-vbs}.
%


This motivates \sys to partition the responsibility of anti-cheat enforcement along the host and guest boundaries (\autoref{t:primitive-vbs-repartition}).

\begin{packeditemize}                                                                 
    \item {\bf System-level enforcement}, including boot integrity, driver verification, anti-debugging, device isolation, and remote attestation, are generic platform security features that PVM hypervisors enforce.
    The developer simply attests to these features when a player wants to start a new game session.
    While the game session is in-play, the hypervisor ensures that the attested state is not modified (e.g., prevents new driver loading).

  \item {\bf Game-specific enforcement}, including memory scanning, binary hardening, 
  asset and module integrity, and game data collection, are enabled by user-space anti-cheat components that developers already ship alongside today's game clients 
  (\autoref{s:kernel-level-anticheat}).
  These components can operate directly within the protected guest environment alongside the game binary.
  %
\end{packeditemize}

\subsection{\Revision{(R4)}{Deployment Assumptions}}

\Revision{(R4)}{The fragmentation of the Linux ecosystem poses a remote attestation challenge because users and distribution maintainers reserve the right to customize trusted components and kernels. 
Therefore, our immediate deployment target are mature distributions (e.g., Ubuntu, SteamOS) and controlled platforms (e.g., Steam Deck, Steam Box). 
In these targets, maintainers can actively track the platform status like Microsoft does for Windows setups~\cite{microsoft2025healthattestation}.
Gaining enough traction, we believe the Linux community will standardize remote attestation requirements.
A precedent for this can be found in prior standardization work for UEFI Secure Boot~\cite{burke2012uefi}.
}

\section{System and Threat Model}
\label{s:assumptions}

%


The game player has root (or admin) privileges on their computer and the technical ability to launch software-based attacks against video games (e.g., wallhacks via memory tampering). 
The player does not trust game-developer code, which may attempt to disclose host data, while the developer does not trust the player.
%

Both parties trust the OS vendor (e.g., the Linux distribution provider) to ship the
platform.
From the game developer's perspective, the host kernel is initially benign but, since the player has administrative control, may be manipulated at runtime (e.g., by installing malicious drivers).
This assumption is exactly the one modern OSes already adopt: Windows VBS and Android
Protected KVM are deployed precisely because the OS no longer trusts the kernel to be vulnerability-free and instead relies on hardware-assisted isolation and integrity hardening.
Within our architecture, the hypervisor is the trusted computing base (TCB) on which both parties rely for isolation, integrity, and mediation across the host and game
environments.
The hypervisor is launched before the host OS through secure boot~\cite{sysguard},
becomes permanently resident, extends its measurements into the TPM, alongside the host's initial state and security configuration.
All parties can query the TPM to attest that the system is running atop a trusted hypervisor.                     

\PP{Out-of-scope}
Consistent with current kernel-level anti-cheats, we exclude hardware-based attacks that rely on the interposition of display or input devices~\cite{s4dbrd-anti-cheat}.
These attacks are generally expensive and error-prone.
We do not consider memory-based attacks (e.g., control-flow hijacking through ROP/BROP) that allow attackers to compromise the kernel without installing malicious drivers.
Finally, we also consider micro-architectural defects~\cite{plundervolt} and side-channels~\cite{spectre,meltdown} as out-of-scope.
\AD{Anything else?}

%% file: tables/primitives-VBS.tex
\begin{table}[t]
\caption{
Typical protection primitives (P0 - P9) required by today's game protection software~\cite{eac,riot_vanguard_2024,Dorner2024examination}.
}
\label{t:primitive-vbs-repartition}
\centering
\footnotesize
\begin{tabular}{l}
\toprule
{\bf System-Level}~({\em implemented by Kernel drivers}) \\
~~{\bf P0:} Block unsigned/malicious kernel drivers from loading at boot. \\
~~{\bf P1:} Block debuggers/inspectors that read or modify the game process. \\
~~{\bf P2:} Block external devices from accessing memory or synthesizing input. \\
~~{\bf P3:} Restrict runtime driver loading to signed and known-good drivers. \\
~~{\bf P4:} Remotely attest platform integrity to the game server. \\
\midrule
{\bf Game-Specific}~({\em implemented by userspace component}) \\
~~{\bf P5:} Scan game memory for known cheat signatures/suspicious writes. \\
~~{\bf P6:} Obfuscate code and data to slow reverse engineering. \\
~~{\bf P7:} Detect unauthorized modifications to game binary and asset files. \\
~~{\bf P8:} Detect or block unauthorized DLL/module injection to game. \\
~~{\bf P9:}  Collect game runtime data (screenshots, player events). \\
\bottomrule
\end{tabular}
\end{table}

%% file: background.tex
\section{Running Video Games under Virtualization}
\label{s:virtualization}

\AD{TODO: Adil}

Our proposed approach~(\autoref{s:overview}) requires video games to execute within virtualized environments.
This section investigates the suitability of the current virtualization stack for this approach.
It begins by providing preliminary knowledge related to virtualization and video game requirements, and then discusses our findings based on a real-world case-study of running games in virtual machines.
\subsection{Preliminary Knowledge}
\label{s:study:preliminary}

Modern computers can leverage hardware support~\cite{intel-manual} to virtualize CPUs at near-native speeds, but share devices between a guest and host using slower software techniques~\cite{vm-game}.
In terms of video games, the most latency-sensitive device is the Graphics Processing Unit (GPU).
%
The remaining paragraphs explain the pipeline (in Linux-based systems) through which games access the GPU in native (non-virtualized) and virtualized environments.
%


\begin{figure}[t]
    \footnotesize
    \begin{center}
      \centering
    \includegraphics[width=0.9\columnwidth]{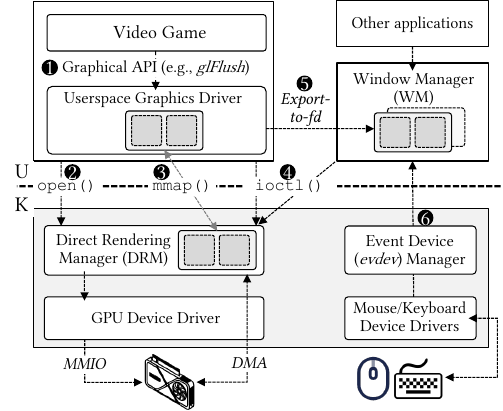}
      \caption {Native graphics rendering and display pipeline.}
     \label{f:mesa-init}
    \end{center}
\end{figure}

\PP{{\em Native} graphics processing pipeline.}
Video games coordinate with the GPU device to render frames (i.e., generate images from game state or 3D models) and ask the Window Manager (WM) to display these frames~(\autoref{f:mesa-init}).
%
%


%
%

\vspace{0.2em}
\noindent
{\em Graphics rendering via user-space and kernel drivers.}
The game issues render (or draw) calls through a standardized graphics API, such as OpenGL or Vulkan on Linux~(\BC{1}).
These APIs are handled by a user-space graphics driver (or library) that is included within the game process.
%
%
The library interacts with the GPU through
%
the kernel's {\em Direct Rendering Manager} (DRM) subsystem~\cite{linux-drm-internals}.
Like the user-space driver, 
the role of DRM is to maintain a standardized interface to send data and commands to underlying GPU device driver (and hardware).
It exposes a virtual render node (e.g., \cc{/dev/dri/128}) corresponding to a GPU device that the user-space driver opens for access~(\BC{2}).
The user-space driver then interacts with DRM along two complementary pathways:
\begin{packeditemize}
    \item {\em Data pathway via memory-mapped GEMs~(\BC{3}):}
    DRM allocates {\em Graphics Execution Manager} (GEM) objects to hold rendering data such as framebuffers, textures, and vertices; the user-space driver \cc{mmap}s them into its own address space and writes directly, avoiding per-frame memory copies.

    \item {\em Command pathway via system calls~(\BC{4}):}
    With data staged in GEMs, the driver submits commands and synchronization primitives through \cc{ioctl}; DRM forwards them to the GPU device driver, which programs the GPU over MMIO while bulk data moves through Direct Memory Access (DMA).
\end{packeditemize}

\vspace{0.1em}
\noindent
{\em Graphics display via the window manager.}
Display (to the screen) is handled by a system-wide window manager daemon, such as an X11 server or a Wayland compositor~\cite{wayland_arch}.
Once a frame has been rendered, DRM's \cc{export-to-fd} capability hands the GEM handle directly to the WM, which asks the GPU driver to scan the frame out without data copies between different userspace contexts~(\BC{5})~\cite{wayland_arch}.
%
%
The WM is a higher-privileged user-space daemon that composes surfaces from all running applications at their respective resolutions and scaling.
%
Note that the window manager also facilitates player input (e.g., mouse and keyboard).
%
The player input events happen and enter the kernel through their device drivers, are normalized into a uniform stream by the Event Device (\cc{evdev}) subsystem, and are forwarded by the WM to the focused application (the game) using inter-process communication~(\BC{6})~\cite{linux_input_subsystem, wayland_arch}.

\begin{figure}[t]
    \footnotesize
    \begin{center}
      \centering
    \includegraphics[width=0.95\columnwidth]{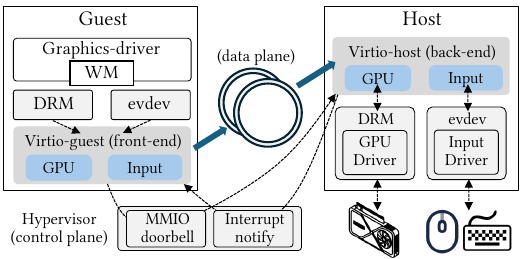}
      \caption {Overview of virtio-based device virtualization.
      }
     \label{f:virtio}
    \end{center}
\end{figure}

\PP{{\em Virtualized} graphics processing pipeline}
The most efficient way for guest to access devices today is to leverage {\em paravirtualization}, an approach where the host provides an emulated device interface through which the guest sends and receives data to/from the device.
The default interface in Linux is called {\em virtio}, which supports shared ring buffer message passing and a doorbell mechanism~(using MMIO) to notify the guest and host.
%
%

\Revision{(R8)}{It is important to note that paravirtualization techniques like {\em virtio} are unsafe in general trusted execution environments (e.g., Confidential VMs like Intel TDX~\cite{tdx} running in cloud machines) because an administrator can read or tamper with the communication (e.g., by installing a malicious kernel module). 
In our architecture, a hypervisor implemented by the trusted OS vendor allows attestation and runtime enforcement of the host's integrity state~(\autoref{s:approach:split}). 
Therefore, after attestation, paravirtualization is suitable in our setting since the attested and protected host kernel will not allow an administrator to hook onto VM-device communication.}

\CQ{
I don't get this paragraph.
I feel it's complicated to mention TDX here. They don't assume a trusted host OS, but we do.
Why not simply say (i) virtio host/guest drivers are attested + (ii) kernel code protection so that no one else can manipulate the virtual device channel?
}

%
\autoref{f:virtio} illustrates paravirtualization for GPUs in Linux.
The {\em virtio-gpu} device allows sharing of graphics data to the host.
%
A thin guest-side driver surfaces a virtual GPU device to the game's userspace graphics stack and the kernel's DRM subsystem to receive data payloads in batches.
Every batch crosses the guest/host boundary along two coupled planes.
The {\em control-plane} world switches through the hypervisor (an MMIO write in the guest triggers a VM exit, the hypervisor dispatches to the host back-end to handle the commands, and an interrupt injection returns the result).
The commands are followed by a {\em data-plane} copy through the shared ring
carrying the per-batch graphical command payload.
The device offers two ways to carry payloads using a higher-level GL/Vulkan command stream (so-called VirGL) or a lower-level vendor DRM stream (so-called DRM native context):

\begin{packeditemize}

    \item {\bf VirGL~\cite{virgl_mesa}:}
    The guest encodes each graphical API call (e.g., OpenGL) into a portable, vendor-independent command stream and sends it to the host; the host decodes the stream and re-issues the calls against the real GPU's user-space driver.
    The encode/decode step makes VirGL portable, but adds a per-call translation cost on both sides of the boundary.

    \item {\bf DRM native context~\cite{virtgpu_native_context}:}
    Instead of encoding/decoding graphical API calls, the game's user-space graphics driver is connected directly to the host's DRM subsystem.
    This allows the application to issue the same per-vendor \cc{ioctl}s and GEM operations it would on bare metal, while {\em virtio-gpu} acts as a thin transport for those calls~(e.g., copying data between GEM).
    %
    
    %
\end{packeditemize}

%% file: design.tex
\subsection{Case-Study and Findings}
\label{s:study:findings}

We conducted a case study with the state-of-the-art virtualization stack for gaming in Linux systems.
Concretely, we ran {\em muvm}~\cite{muvm_github}, a lightweight gaming-focused VM that runs a cut-down version of Arch Linux on the optimized {\em libkrun} virtual machine manager~\cite{libkrun}.
{\em muvm} leverages hardware acceleration through KVM and supports paravirtualization~(\autoref{s:study:preliminary}), both VirGL and DRM native context.
Inside the guest, we ran three popular open-source Linux games and compared their performance to execution on the host.
Refer to~\autoref{s:perf} for details about the hardware and games.
%


%
Even before starting a game, we made the observation that this virtual machine setup significantly increased resource utilization compared to native execution.
Linux, by default, is an inherently multi-tasking OS implementation that spins-up multiple background services, system daemons, and auxiliary processes alongside the game task.
Many of these tasks are not required in our setting, but it is challenging to disable them entirely from the kernel.
The result of this was that in one game the CPU utilization of the system went by $4.9\times$, while the system also consumed approximately $600$MiB of additional memory for running the same game.
The memory impact, while small for dedicated gaming machines, would be significant for low-resource computers (e.g., laptops).
More concerning than the resource utilization aspect, is the fact that full kernel implementations like Linux are complex and huge, making it challenging to reason about isolation between the guest and host environments.
Specifically, recent work~\cite{gramine-tdx} has analyzed the interfaces between lightweight Linux guest kernels~\cite{firecracker} and the hypervisor/VMMs, estimating that around 911 such pathways for communication exist.
Validating each of these pathways to ensure host privacy as well as guest protection is a non-trivial task.
%



\begin{figure}[t]
    \footnotesize
    \begin{center}
      \centering
    \includegraphics[width=0.98\columnwidth]{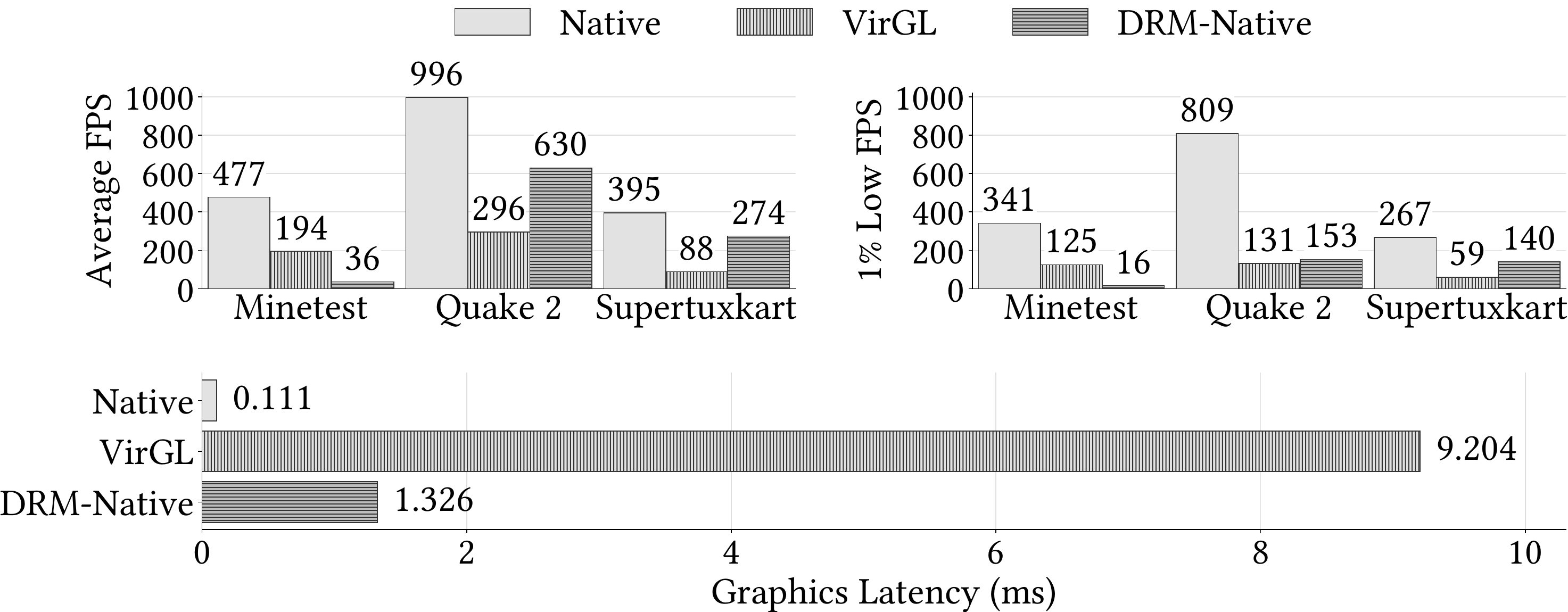}
      \caption {The top graphs show performance for existing graphics solutions across three games. 
      The bottom graph shows graphics pipeline latency in SuperTuxKart at 1080p.}
     \label{f:study-breakdown}
    \end{center}
\end{figure}


\begin{mybox}
{\bf Observation-1:} {\em Full-fledged OS kernels raise both resource utilization and isolation challenges in the gaming context.}
\end{mybox}



We next examined graphical performance using the end-to-end {\em frames-per-second (fps)}, the standard metric for real-time gameplay, within the guest environment.
\autoref{f:study-breakdown}~(top graphs) reports the average and $1\%$-lows fps inside the guest.
We noticed that the games produced $1.4$-$13.3\times$ and $1.9$-$21.3\times$ fewer frames, respectively, against native execution, well below what high-performance games today tolerate.
This overhead was felt for both VirGL and DRM native context~(shown as DRM-Native in the figure).
%

To localize the cause of this performance drop compared to native execution, we measured the roundtrip time on each graphics command submitted to DRM~(at \cc{ioctl}) and on each frame submitted to the WM~(at IPC) in both the host and guest environments.
\autoref{f:study-breakdown} (bottom) illustrates the average graphical pipeline latency for one game, and we note that both paravirtualization solutions incurred at least an order of magnitude overhead in terms of their graphics pipelines.
In the worst-case, VirGL's graphical pipeline was upto $82.9\times$ slower than native execution.
Note that this latency is amortized by CPU computation time (e.g., executing game logic and preparing frames to be rendered), which is roughly the same inside the guest and host due to efficient hardware CPU virtualization, and that is why the impact on {\em fps} is marginally lower.
%

%
The increased graphical pipeline latency is not surprising looking at the guest-host crossing of {\em virtio-gpu}~(\autoref{s:study:preliminary}) paid on every batch job submission. 
Recall that this incurs an MMIO-triggered guest exit, a hypervisor dispatch to the host backend, a shared-ring copy of the per-batch payload, and an interrupt-injection return.
VirGL additionally adds a per-call encode on the guest and a matching decode on the host.
DRM native context removes the translation step (and does improve performance in most scenarios compared to VirGL) but leaves the per-batch world switch intact.
%


\begin{mybox}
{\bf Observation-2:} {\em Virtio-based GPU paravirtualization pathways impose excessive cost in intensive video game scenarios.}
\end{mybox}

\section{\sys Design}

\sys is a privacy-friendly gaming architecture that is built on our proposed approach of virtualization-based sandboxing and split responsibility architecture~(\autoref{s:overview}).
To address the limitations of the current virtualization software stack in terms of supporting our approach~(previous section), this section introduces two new design ideas: 
(a) a Library OS kernel design that is lightweight and minimized in terms of its outside interfaces specifically to support Linux games~(\autoref{s:design:libos})
and 
(b) a novel graphics sharing pipeline for our proposed kernel design that does not rely on slow paravirtualization interfaces provided by existing solutions~(\autoref{s:design:rendering}).
%


%

\subsection{Minimized Game-Centric LibOS Kernel}
\label{s:design:libos}

A Library Operating System (LibOS) kernel provides a specialized execution environment within guest virtual machines to support a single computation.
Like traditional kernels, LibOS kernels contain several subsystems that ensure smooth execution of the application inside the guest including system call handlers, scheduling, memory management, socket handling, and device drivers.
Additionally, modern LibOS kernels provide wide support for the POSIX system call abstraction, which is critical for running unmodified Linux applications.
Famous examples of LibOS kernels include Gramine~\cite{gramine-tdx} and UniKraft~\cite{unikraft}.
%

%

\begin{figure}[t]
    \footnotesize
    \begin{center}
      \centering
    \includegraphics[width=0.85\columnwidth]{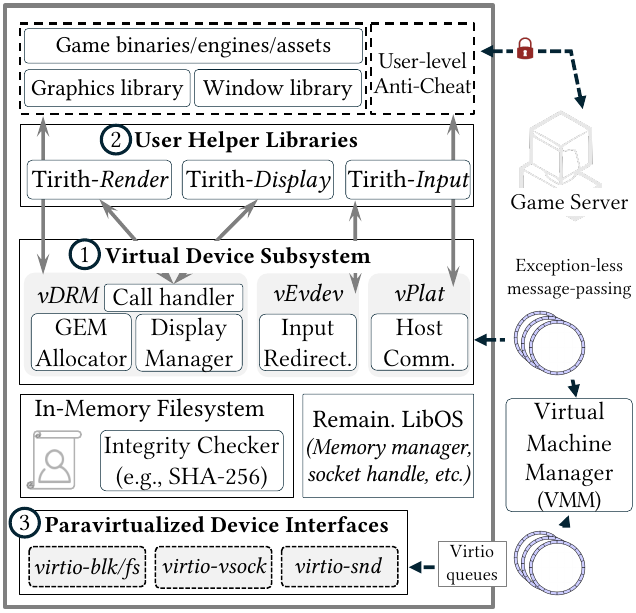}
      \caption {\sys LibOS architecture and components.\AD{TODO: remove/minimize whitespace and shrink. make more precise as well. Add sound}}
     \label{f:libos}
    \end{center}
\end{figure}

%
There are two problems in leveraging existing LibOS implementations for \sys. 
First, existing LibOSs are designed for cloud computations and they lack support for devices like graphics and input.
A naive approach would be to port the infrastructure for these devices from Linux kernels, but this is complex given the differences between Linux and LibOSs, as well as potentially unsafe (i.e., opens unnecessary attack surfaces).
Second, for the same reason as above, existing implementations are neither designed for game delivery nor coordination with the (trusted) hypervisor for split game protection.
This section describes how we address both problems. 
\autoref{f:libos} illustrates the overall LibOS architecture.
%

%

\PP{Game-specific device interfaces and drivers.}
Recall that Linux games interact with critical devices using the virtual device subsystem~(\autoref{s:study:preliminary}),
thus \sys implements a virtual device subsystem that supports backward-compatibility of existing games.
Also, \sys provides other paravirtualized device interfaces. 
Both the subsystem and interfaces are designed to maintain only a small set of controlled, outside interfaces~(refer to~\autoref{s:security-analysis} for details).
%


\vspace{0.2em}
\noindent
\WC{1}~{\em Virtual Device Subsystem:}
\sys's device subsystem implements four devices, including a virtual Direct Rendering Manager~(DRM) for graphics, event device for input, sound device, and a platform device for host state attestation.
The main role of these subsystems is to intercept system calls from user space and redirect them either to internal \sys-LibOS components (e.g., helper libraries~\WC{2}) or the host virtual machine manager.
Since communication to these devices is latency-sensitive, \sys implements an efficient exception-less message-passing interface with the host. 
We explain this message-passing interface and the function of the graphics and input subsystems in the next section~(\autoref{s:design:rendering}).
The platform device facilitates coordination between the hypervisor and the guest code to establish and maintain the trustworthiness of the host using three steps.
First, it allows the guest to query platform state and measurement related to boot events~(e.g., stored in the computer's TPM). 
Second, it allows the guest to obtain a signed attestation digest that can be reported to the remote game server for verification.
Third, it allows the game developer to determine the currently-loaded kernel drivers and ask the hypervisor to prevent further loading of drivers until the game is terminated.
Each of these aspects are enabled by calling the trusted hypervisor (using hypercalls).
Note that game developers can add more features to implement more robust game protection in the future.
%



\vspace{0.2em}
\noindent
\WC{3}~{\em Paravirtualized Device Interfaces:}
While our LibOS implements new solutions to access GPUs and input devices, it leverages {\em virtio} to share other~(less latency-critical) devices with the host.
Specifically, games require access to (a) the network to send/receive information from game servers, (b) the sound card for audio playback and microphone capture, and (c) storage to retrieve files and game assets.
The LibOS relies on {\em virtio-vsock} and {\em virtio-snd} to support network and sound, respectively.
The virtual socket~({\em vsock}) interface is optimized for direct socket interactions (i.e., it does not require expensive operations like virtio-gl), and it avoids an implementation of the complex network stack inside the LibOS.

For storage, \sys supports two interfaces, namely {\em virtio-blk} and {\em virtio-fs}~\cite{russell2008virtio, stefanha2019virtiofs}.
The former is a high-performance interface that allows loading entire virtual machine images~(as block devices) into the guest environment.
\sys leverages this interface to efficiently load the initial game image~(next heading).
The latter interface allows reading and writing to files on the host. It is used at runtime to read additional files if needed, or write-back data to the host (e.g., log files).
All files that are read from the device are integrity-checked through the LibOS's in-memory file system using a signed manifest securely provided at runtime~(\WC{3}).
We explain game deployment and loading in the next heading.
\AD{needs a pass.}

%



%



\PP{Game deployment and runtime protection workflow}
This section explains the end-to-end event workflow for executing games with \sys in Protected Virtual Machines~(PVMs).
To support the deployment of video games, \sys provides a template guest image (which we call a {\em Zygote}) to developers.
In addition to previously-described LibOS kernel components, this image contains core helper libraries responsible for implementing efficient graphics and input pipelines~(described in the next sections).
%
%

During offline preparation, the game developer creates a cryptographically signed {\em manifest}, a Zygote, and a game (storage) drive for their particular game.
The storage drive contains the game's primary assets (e.g., game binary and libraries) and its user-level anti-cheat component~(\autoref{s:kernel-level-anticheat}).
The manifest contains integrity hash measurements of the game drive and other file assets that will be loaded into the virtual machine at runtime.
It is signed using the developer's private key.
At this stage, the game developer also installs their public key into their game's Zygote.
This allows loading of correctly-signed game manifests, once the Zygote executes inside the PVM.
The signed manifest, Zygote, and game drive are sent to the user's computer (e.g., through standard game delivery platforms like Steam or Epic Game Library).
%



To start a game, the host invokes the trusted hypervisor to instantiate a PVM using the provided Zygote image.
The hypervisor measures the Zygote at launch time, records these measurements (e.g., in its protected memory regions or a TPM), and enforces strict isolation between the PVM and the host environment.
After boot, the Zygote's LibOS first reads the signed manifest (through {\em virtio-fs}) and verifies it using the installed key.
If the signature passes, it loads the game drive and other files mentioned in the manifest and verifies its integrity using the provided hash.
At this stage, if any file is found to be corrupted the PVM stops execution, since this signals a potentially-malicious host or some other corruption.
Note that game asset integrity-checking is also a common required step by today's anti-cheat software~\cite{epic_anticheat_integrity_tool_config}.
%


After the game environment is set-up inside the PVM, the game connects over the network (through {\em virtio-vsock}) to authenticate itself with the game server.
Specifically, using \sys's platform device, the game obtains the platform's measurement (recorded during secure boot within the host TPM), as well as the measurement pertaining to the initially-loaded Zygote image.
These results are embedded into an attestation digest and sent to the game server.
Game developers can remotely attest the PVM by validating the digest measurements, thereby confirming that the game is executing atop a \sys-enabled hypervisor with an unmodified Zygote and intact protection mechanisms.
Only after successful attestation is the game allowed to start a competitive session.\AD{read}
%




%



\subsection{Shared GEM Context Graphics Pipeline}
\label{s:design:rendering}

\sys implements an efficient graphics pipeline that video games use during execution to avoid the expensive data-plane and control-plane operations of {\em virtio-gpu}~(\autoref{s:study:findings}).
\Revision{(R8)}{ 
The key insight for this pipeline is that instead of sending data frames to the host, we can securely map the host's graphics data-containing structures, namely the Graphics Execution Manager (GEM) objects into the guest and allow it to operate directly on these objects.
This shared mapping is safe, because \sys's anti-cheat architecture can protect the confidentiality and integrity of GEM data regions~(\autoref{s:study:preliminary}). 
In particular, the hypervisor's attestation establishes the trustworthyness of the PVM and host (as described in the previous section), which in turn prevents users from accessing game-related GEMs.} 

\begin{figure}[t]
    \footnotesize
    \begin{center}
      \centering
    \includegraphics[width=0.99\columnwidth]{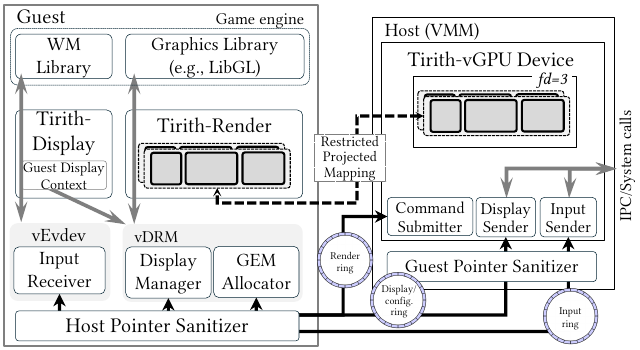}
      \caption {Illustration of Shared GEM context graphics.}
     \label{f:rendering}
    \end{center}
\end{figure}

With the aforemention approach, only command submission must be redirected to the host, which can be efficiently achieved given the small size of control packets and \sys's exception-less message passing, as well as restricted address space projection.
Another observation we make is that, with our new pipeline, rendered frames are already available within the host virtual machine manager (i.e., within the GEM objects).
Therefore, the rendered frames can be {\em directly} displayed by instantiating a window on the host and thus achieve native display performance~(\autoref{f:rendering}).
%


\PP{Cross-Context GEM allocation and mapping}
While GEMs belonging to the host DRM interface and GPU cannot be allocated by the guest, we observe that they can (a) first be mapped into the host virtual machine manager (VMM)'s address space and then (b) be projected into the guest's allocated physical regions.
Recall that GEM objects are unique per-process~(\autoref{s:study:preliminary}), and thus there is no security or privacy problem in projecting the VMM's  GEM objects into the guest.
Moreover, GEM objects only contain {\em static} data (i.e., no internal nested pointers), thus only projecting the GEM regions is sufficient.
The paragraphs below explain how \sys achieves the aforementioned mapping and projection.\AD{check}
%



%
During boot, \sys reserves a small pre-defined region of its allocated physical memory to hold GEM objects.
In our experiments, a region of 128MiB was mostly sufficient for our evaluated applications.~\Revision{(R1)}{
%
In the scenario where the pre-allocated guest region for GEMs is exhausted, \sys leverages the following workflow.
%
%
The game's LibOS kernel first reserves another free physical region.
Then, it notifies the trusted hypervisor of the newly-reserved region's parameters (i.e., address and size) through a secure hypercall.
This allows the hypervisor to track that this region is free and can be re-allocated for GEMs if requested (see below).
%
}
%

%
When the game executes, the userspace graphics library (e.g., OpenGL) requests creation of GEM objects to \sys's {\em vDRM}.
These calls are forwarded to the host VMM~(using the channel described in the next heading).
The VMM asks the host DRM to allocate GEM objects as required into its address space.
The physical frames containing these objects must be mapped into the guest, but the host cannot directly change a PVM's physical regions for security purposes.
Specifically, the Extended Page Tables (EPT) that control guest mappings are controlled by the trusted hypervisor.
Therefore, the VMM sends the physical address of the objects to the guest {\em vDRM}, which first validates these mappings (i.e., they do not belong to a pre-existing guest region).
If validation succeeds, it requests the trusted hypervisor to update mappings.

\Revision{(R2)}{Note that GEM objects contain VRAM-resident buffers (i.e., memory regions directly shared between the GPU and CPU). 
Once these objects are mapped into the guest's address space using the workflow described in the previous paragraph, we must ensure that writes made to these regions by the guest remain visible to the GPU. 
In non-virtualized settings, the kernel automatically handles this by setting a {\em Write-Combining} (WC) policy in page tables related to GEM pages. 
This avoids storing writes to GEM regions in the CPU cache using efficient batching, ensuring cache-coherency across CPU-GPU contexts.
By default, our remapped GEM objects, however, are marked cache-coherent only within the guest context.
}
This means that the guest's write operations on GEM objects may be cached by the CPU, and not be directly visible on the host (and GPU) at command submission time.
\Revision{(R2)}{And, this is critical for VRAM resident buffers, where signaling mechanisms goes over GEMs.}
To address this problem, \sys ensures {\em Consistent Caching Parameters} for the GEM reserved region. 
Specifically, \sys modifies guest page tables after GEM remapping and marks the GEM-related physical pages as Write-Combine~(WC).
Note that this policy is only required for the GEM reserved region.
\PP{Control redirection for render command submission}
With the VMM's graphical data-plane projected directly into the guest, \sys must ensure that control commands are appropriately forwarded to the host for submission.
Recall that this includes system calls like \cc{ioctl} for rendering~(\autoref{s:study:preliminary}).
\Revision{(R2)}{Additionally, \cc{ioctl} is also used for CPU-GPU synchronization in Linux.
In particular, applications use these calls to both allocate kernel-backed fences (\cc{DRM\_IOCTL\_SYNCOBJ\_CREATE} parameter) and signal the GPU (e.g., \cc{DRM\_SYNCOBJ\_WAIT}) using the DRM as an intermediary.
}

A naive way would be to leverage a {\em virtio} interface to redirect control commands, but it incurs significant latency \Revision{(R2)}{that is particularly harmful for frequent synchronization operations}. 
\sys avoids this by implementing an exception-less message-passing interface~\cite{scone,flexsc}.
This interface leverages shared memory between two communicating contexts and background threads that are quickly woken-up on message arrival.
Specifically, \sys's interface contains two shared regions, including a {COMM} and {DATA} region, and uses CPU notifiers to wake-up threads.
Our evaluation indicates that our implementation is efficient even under a small number of available system threads.
%



%

While other control-plane commands are self-contained, \cc{ioctl} packets raise a complication because they may internally contain pointers to other in-memory data structures in guest memory~\AD{cite}.
For example, take a given \cc{ioctl} which takes as an argument, a reference to a struct that contains various other objects which may be located on the stack, heap, or other memory segments.
One way to address this problem is to copy/translate pointers across contexts~\cite{ksplit}, but this would incur extra costs and reliability problems.
%

%

%
To address the aforementioned complication, \sys implements {\em Read-Only Identity-Mapping} of the guest's data regions into the host virtual machine manager.
%
Specifically, the guest's data regions are mapped at the same virtual address in the VMM, allowing both the guest and host to reference points using the same address,  removing the complexity and wasted computation of translating object references at runtime.
Note that this process is simplified in a LibOS context, because (a) LibOS kernels have a flat address space where the virtual and physical address is identical and (b) the heap, global, and data regions are determined during boot~\cite{gramine-tdx}.
\sys achieves identity-mapping by initially reserving a part of the VMM's address space before the PVM launches, and then requesting the trusted hypervisor to map the physical frames of the guest to the reserved space.

Sharing direct pointers across untrusted contexts can result in confused deputy attacks (e.g., the guest passes a malicious pointer that belongs to the host's memory to steal data~\cite{boomerang}). 
%
%
\sys addresses this problem by implementing {\em Pointer Sanitization} at both the guest and host entry-points by checking that pointer addresses are valid (i.e., guest pointer only points to guest regions and vice versa) before relaying any command to internal components.  
\PP{Direct display with bridged windowing}
Due to shared GEM objects, rendered frames are already available within the host context.
%
%
To avoid cost for displaying these frames, \sys creates a window on the host that fully mimics the game's intended operations (e.g., resolution, etc.) and directly displays the frames on this window.
Specifically, \sys's display library intercepts all windowing-related calls from the game engine and presents to it a virtual display context inside the LibOS environment.
For each of these commands, it coordinates with the host VMM to perform these operations as required~(explained below).
%
%
%

%
There are three main classes of windowing-related calls: (a) configuration calls which change display options like create windows and set resolutions~(e.g., \cc{XCreateDisplay} for X11 windows), (b) display calls that send frames to be displayed and synchronize their execution, and (c) input calls that receive player interactions from keyboard, mouse, and other devices. 
The first two classes of calls must be sent from the guest to the host, while the third class must be sent from the host to guest.
%


%
\sys leverages two exception-less message-passing interfaces to handle these calls~(i.e., one for the guest-to-host channel and one for the host-to-guest channel).
In particular, all input-receiving calls~(e.g., \cc{XSelectInput}) are intercepted at the host library and relayed to the guest using shared memory.
Note that since input latency is critical for many video games, \sys creates a dedicated channel (with its own background threads) just for input redirection to minimize latency~(refer~\autoref{s:perf:benchmarks} for details).
The remaining calls are intercepted at the guest and relayed to the host using the same exception-less memory-based communication channel established in the previous section.
%

%

Like graphical rendering commands, window-related calls also contain nested pointers.
Recall that guest data regions are already shadowed in the host's address space~(previous heading), thus guest-to-host calls and embedded pointers are automatically handled.
For host-to-guest pointers, \sys creates a reserved heap region within the host VMM, which is similarly shadowed into the guest's address space.
The host leverages a custom memory allocator on this region and receives input commands from the host on this region.
This allows {\em zero-copy} message-passing both ways.
As we explained previously, pointers are always checked when passing between guest-host contexts to prevent attacks.

%% file: case-study.tex
\input{tables/security-primitives}

%% file: security.tex
\input{tables/outside-interfaces}

\section{Security Analysis}
\label{s:security-analysis}

This sections analyzes \sys's design and architecture to show that it provides both player host privacy and an equivalent level of game anti-cheat protection (as kernel-level anti-cheat).
Note that \sys's virtualization-based security monitor is the first component loaded during system boot and serves as the root of trust.
%
%
%
%

\PP{Newly introduced interfaces.}
Beyond the monitor itself, all interaction between \sysgame and the host traverses a small, auditable set of interfaces, summarized in \autoref{t:outside-interfaces:privacy}.
Following the discipline of recent confidential-VM frontends~\cite{gramine-tdx}, we keep this surface deliberately narrow: each interface has a fixed direction, a single mediator, and carries only a well-typed payload.
Compared to native Gramine-VM~\cite{gramine-tdx}, which already exposes virtio-blk/-fs for storage and virtio-vsock for network, \sys introduces four new interfaces (I1--I4) for graphics, windowing, input, and host attestation.
The two subsections that follow analyze the host-privacy and game-security guarantees.
%

\AD{21 existing, we add 5. Way less than Linux (e.g., 911 as shown by prior work~[yyy])}

\subsection{Host Privacy Protection}
\label{s:security:privacy}
The game developer's anti-cheat executes with full kernel privileges inside the guest.
A malicious developer could attempt to extract data from the host by:

\begin{packeditemize}
\item {\em Directly reading host memory of processes or devices or issuing DMA from a guest-controllable device;}
\end{packeditemize}

The guest executes within a Protected VM~(PVM) whose CPU state and memory are isolated from the host by the hypervisor's Extended Page Tables, and whose DMA-capable devices are constrained by the IOMMU and interrupt-remapping.
Specifically, the guest cannot leverage the CPU to read or write to any memory region that is not shared with it. Moreover, the guest does not have direct access to any physical device to launch DMA attacks; all of its devices are paravirtualized and hence mediated~(next heading).
%

\begin{packeditemize}
\item {\em Abusing the in-band interfaces of \autoref{t:outside-interfaces:privacy} to leak host state or perform unsanctioned host actions.}
\end{packeditemize}
{Each interface in \autoref{t:outside-interfaces:privacy} is constrained by its mediation checks so that it cannot leak unrelated host state or coerce unsanctioned host actions.}
In particular, the guest can try to leak host data using the message-passing interfaces~(I1) (e.g., Boomerang~\cite{boomerang}). 
The host-side sanitizer bounds-checks every pointer against the guest's allocated regions to block these attacks.
For the bridged windowing calls (I2), the bridge maintains an allow-list of lifecycle API (create/resize/destroy/swap) operating on game's own window: 
for instance, the bridge refuses any cross-window or top-level overlay operation that would let the game to draw on or read from other host windows and break user privacy.
For the shared GEM regions (I3), the trusted hypervisor projects only host-allocated GEM objects into a pre-reserved guest physical region; the guest cannot enlarge this mapping or redirect it to other host regions.
Last, for the trusted-monitor hypercall (I4), the hypervisor exposes only an allowlisted set of platform-measurement, driver-enumeration, policy-tightening, and attestation-digest queries; it returns measurement results rather than raw host state and never executes developer-supplied code on the host.
%

%


\subsection{Game Security Enforcement}
\label{s:security:game}
The player retains administrative privileges on the host, but remains constrained by the trusted hypervisor from gaining malicious access to game state.
%
%
%
A malicious player can try to do this by:

\begin{packeditemize}
\item {\em Reading or modifying the guest's memory or devices using host privilege (e.g., debuggers, drivers, or DMA devices);}
\end{packeditemize}

Like in the opposite direction, the hypervisor's memory isolation prevents the host from mapping the guest's pages, debugging across the boundary, or DMAing into guest memory.

\begin{packeditemize}
\item {\em Tampering with the host kernel and modifying important kernel functionality (e.g., using malicious drivers);}
\end{packeditemize}

The hypervisor also measures the initial state of the host kernel (i.e., during secure boot) and tracks which drivers are being executed on the kernel. 
Therefore, if the user changes system configurations or installs malicious drivers, it will be reported~(\autoref{s:design:libos}).


\begin{packeditemize}
\item {\em Abusing file system access to load malicious files and network interfaces to tamper with network packets;}
\end{packeditemize}

Each file loaded into the guest is integrity-checked using an integrity hash contained within the signed manifest provided by the developer, catching any tampering attempts.
Furthermore, network communication is end-to-end encrypted between the guest and the game server using remote attestation and TLS connections.
%


\begin{packeditemize}

\item {\em Booting without the trusted hypervisor;}
\end{packeditemize}

Remote attestation of the TPM-anchored boot measurement lets the game server detect a non-virtualization boot at session start.

\SL{seems that many of them are not essential due to \sys{}? which ones are still required and which others are defense-in-depth?} \CQ{fix} \AD{took a small pass.}

\subsection{Cheat Prevention Case Studies}
\label{s:case-study}

\sys preserves every enforcement category of today's kernel-level anti-cheats~(\autoref{t:primitive-vbs-repartition}).
%
The system-level enforcement is delegated to the trusted hypervisor and attested by the game server at session start.
%
The game-specific enforcement is, as in today's anti-cheats, already implemented as components packaged with the game client (\autoref{s:kernel-level-anticheat}); in \sys it ships unchanged into the guest, where the developer holds full authority.
\Revision{(R7)}{
To further illustrate this, we have designed systematic PoC tests that simulate attack behavior and show how \sys enforces protection against three well-known classes of cheats attempted on competitive games.}

\PPn{Wallhacks~\cite{laurens2007novel}.}
These attacks allow players to shoot across walls, and are perpetrated by intercepting the game’s OpenGL library
(e.g., DLL injection) to tamper with depth-testing operations.
With \sys, the library executes inside the protected guest which cannot be debugged~(C3). Only developer-specified  modules and libraries can be loaded into the guest~(enforced through the manifest), and these are verified through integrity checks before being enabled.
\Revision{(R7)}{Our PoC test tried to install a custom \cc{.so} file to intercept graphic library calls (OpenGL) through \cc{LD\_PRELOAD}. The LibOS flagged the \cc{.so} file and terminated the guest.}

%
%
%

\PPn{Aimbots and triggerbots~\cite{kanervisto2022gan}.}
Memory-based aimbots reverse-engineer in-memory player coordinates from outside the game to compute aim, or install a host-side hook that injects synthetic mouse/keyboard events. 
Out-of-VM memory access is blocked by \sys; in-guest memory inspection by an injected module is further rejected by the LibOS' FS/module verification.
\Revision{(R7)}{
Our PoC test to simulate this cheat attempts to reverse-engineer in-memory player coordinates from an attack process through \cc{process\_vm\_readv}.
It was unable to read isolated VM memory and thus the attack fails. 
}

It is worth noting that cheaters could leverage visual models (e.g., object-detection on display) and implement visual-aimbots via synthetic inputs.
%
\Revision{(R9)}{Like kernel-level solutions, \sys does not directly address this attack, rather it retains the information that anti-cheats rely on for visual aimbot prevention.
Contemporary anti-cheat systems primarily thwart visual aimbots by detecting unapproved third-party hardware/drivers and applying data-driven heuristic or machine learning analysis to player event trajectories~\cite{steam_vac}. \sys preserves both protection mechanisms. The hypervisor continuously monitors the host to detect and report unauthorized hardware or driver interfaces to the game server.
Furthermore, runtime game data collection, including in-game event logging and screen capture mechanisms, functions without modification inside the PVM, ensuring developers retain full telemetry to analyze player behavior.}

\PPn{Map hacks~\cite{openconflict,kanervisto2022gan}.}
RTS/RPG map hacks lift the fog of war by either correlating in-memory state with network packets or by sniffing the packets through a malicious kernel module on the host network stack (e.g., {\em PUBG-Radar}).
The host's driver-admission policy refuses unsigned modules (can be verified by the trusted hypercalls), foreclosing the standard malicious-driver vector.
Moreover, host network stack would only see packets that are end-to-end encrypted between the game server and the guest.
\Revision{(R7)}{Our PoC test for simulating this cheat relies on having an unsigned kernel module installed at the host to hook host network stack. Our hypervisor detected and flagged this module.
}





%% file: tables/outside-interfaces.tex
\begin{table*}[t]
\caption{
Interfaces newly introduced by \sys between PVM and the host.
%
%
``G'': PVM; ``H'': host.
}
\label{t:outside-interfaces:privacy}
\centering
\footnotesize
\setlength{\tabcolsep}{4pt}
\renewcommand{\arraystretch}{1.1}
\begin{tabular}{@{}l p{0.14\textwidth} p{0.18\textwidth} p{0.22\textwidth} p{0.34\textwidth}@{}}
\toprule
\textbf{\#} & \textbf{Interface} & \textbf{Carries} & \textbf{Potential Attacks} & \textbf{Mediation and Checks} \\
\midrule
I1 & Exception-less\,message-passing {\scriptsize (G$\to$H, \autoref{s:design:rendering})}
   & Forwarded \cc{ioctl} / graphics syscalls
   & Boomerang-style pointer args reading host memory~\cite{boomerang}
   & Host wrapper bounds-checks pointers against G's identity-mapped data region \\
I2 & Bridged windowing calls\newline {\scriptsize (G$\to$H, \autoref{s:design:rendering})}
   & Lifecycle calls (create / resize / destroy / swap)
   & Screenshot of unrelated host windows~\cite{all-your-pixel}; cross-window overlays
   & Allowlisted lifecycle API on G's own window; no read-back call; no cross-window overlay \\
I3 & Shared GEM data plane\newline {\scriptsize (H$\to$G, \autoref{s:design:rendering})}
   & Per-process host-allocated GPU buffers
   & Projecting attacker-chosen host pages into the guest
   & Hypervisor projects only host-allocated GEMs into a pre-reserved region; not enlargeable \\
I4 & Trusted-monitor hypercalls {\scriptsize (G$\to$H, \autoref{s:design:libos})}
   & Attestation digests, host driver verification, policy enforcement
   & Excessive host introspection or unsanctioned host actions
   & Allowlisted query/policy-tightening set; returns signed measurement digests, not raw host state; no host-side code exec \\
\bottomrule
\end{tabular}
\end{table*}

%% file: implementation.tex
\section{Implementation}
\label{s:impl}


We built a \sys prototype for AMD x86 systems running Linux with the KVM hypervisor backend.
We could not use protected KVM ({\em pKVM})~\cite{pkvm}, which is the new hypervisor backend for supporting Protected VMs, because it is currently under-development~\cite{linux_pkvm_doc} and being ported from the Android (ARM) kernel~\cite{pkvm-android}.
Benchmarks indicate that pKVM incurs a small performance penalty compared to KVM, and mainly during initial boot~\cite{perret2022pkvm}.
Prior work has similarly emulated PVM infrastructure~\cite{omnilog}, given emerging deployment of pKVM.
The rest of this section describes our implementation of the guest LibOS kernel and host virtual machine manager (VMM).

Our LibOS kernel implementation is derived from the industry-deployed Gramine~\cite{gramine-tdx,graphene}~(\Revision{(R10)}{6.2k LoC changes}).
%
Other LibOS-like implementations (e.g., UniKraft~\cite{unikraft}) can be leveraged in the future.
Gramine is already POSIX-compatible and supports over 170 system calls. We added \Revision{(R10)}{13 new system calls}~(e.g., \cc{mremap}) to achieve compatibility with Simple DirectMedia Layer (SDL) library. SDL is used by many Linux native games, therefore our compatibility with SDL minimizes the work needed to support similar games.
%
%

To support our efficient graphics pipeline~(\autoref{s:design:rendering}), 
we wrote two helper libraries that are loaded into the LibOS during boot.
The rendering library currently supports the Open Graphics Library (OpenGL) standard (compliant with version 4.6).
%
%
It intercepts OpenGL context creation (\cc{glCreateContext}) and translates them to the newer EGL backend~\cite{khronos_egl_spec} that is designed to render without display integration.
%
%
%
Another library is responsible for direct display in our pipeline by intercepting userspace relevant OpenGL APIs (e.g., \cc{SwapBuffers}). 
%
%
It also manages input redirection.
%



We built the host-side implementation using QEMU v10.1 as our VMM. 
%
%
We built our {\em virtual GPU (vGPU)} as a new paravirtualized device.
This listener handles all the commands passed from the guest. 
It serves syscalls requests from the guest and executes them via a lockless mechanism to ensure performance in multi-threaded game environments by spawning multiple listener threads as required.
In addition to syscalls commands, the listener module serves requests from the guest to execute display related calls, (e.g., \cc{xcb\_present}, \cc{xcb\_resize}). 
%
We also updated the QEMU memory backend to request projected address space mapping~(\autoref{s:design:rendering}) from the hypervisor when the guest starts.
%
%
Our total implementation for the QEMU-side was approximately $1$k LoC\AD{check}.

%% file: performance.tex
\section{Performance Evaluation}
\label{s:perf}




%
\PP{Setup}
We ran all our experiments on a desktop machine with AMD Ryzen 5600X, 32 GB DDR4 RAM, and an AMD ATI Radeon RX 6950 XT with 16GiB GDDR6. 
The system runs Arch Linux with kernel version 6.19.11. 
All game virtual machines~(including those running \sys and other baselines) were assigned 1 vCPU and 8GiB of memory.
A single vCPU was used due to a current limitation in the Gramine LibOS for AMD CPUs: the kernel is originally designed for Intel CPUs and leverages Intel-specific x2APIC virtualization features.
On AMD CPUs, it reverts to the older xAPIC virtualization which significantly degrades performance for all multi-threaded programs, including ones that do not use graphics.
Our experiments on the graphics pipeline~(outside the guest) and results from prior work~\cite{flexsc,eleos} show that techniques like exception-less message-passing that \sys leverages scale nicely across threads.
%

%



%

\PP{Comparison.}
We compared \sys{} against four main baseline solutions.
First, we implemented an enhanced bare-metal solution ({\bf Native}), which leverages \sys's EGL backend~(\autoref{s:impl}).
We found that this solution had better performance than default game backends in {\em all} scenarios.
\Revision{(R5)}{
Second, we ran the games in a raw bare-metal configuration (\textbf{Unmodified Native}), to show the performance benefits \sys{} gains from the aforementioned EGL backend.}
Third, we ran the default Linux graphical configuration for setting up paravirtualized graphics ({\bf VirGL})~\cite{virgl_mesa}.
Fourth, we ran 
the experimental concurrent work that represents the cutting-edge of graphics virtualization performance ({\bf Native-DRM})~\cite{virtgpu_native_context}. 
VirGL and DRM-Native were executed on the state-of-the-art {\em muvm}.
%
%
For running our games, we leveraged 720p, 1080p, and 1440p resolutions.
%

%


\subsection{Micro-Benchmarks}
\label{s:perf:benchmarks}



\begin{figure}[t]
    \footnotesize
    \begin{center}
      \centering
    \includegraphics[width=\columnwidth]{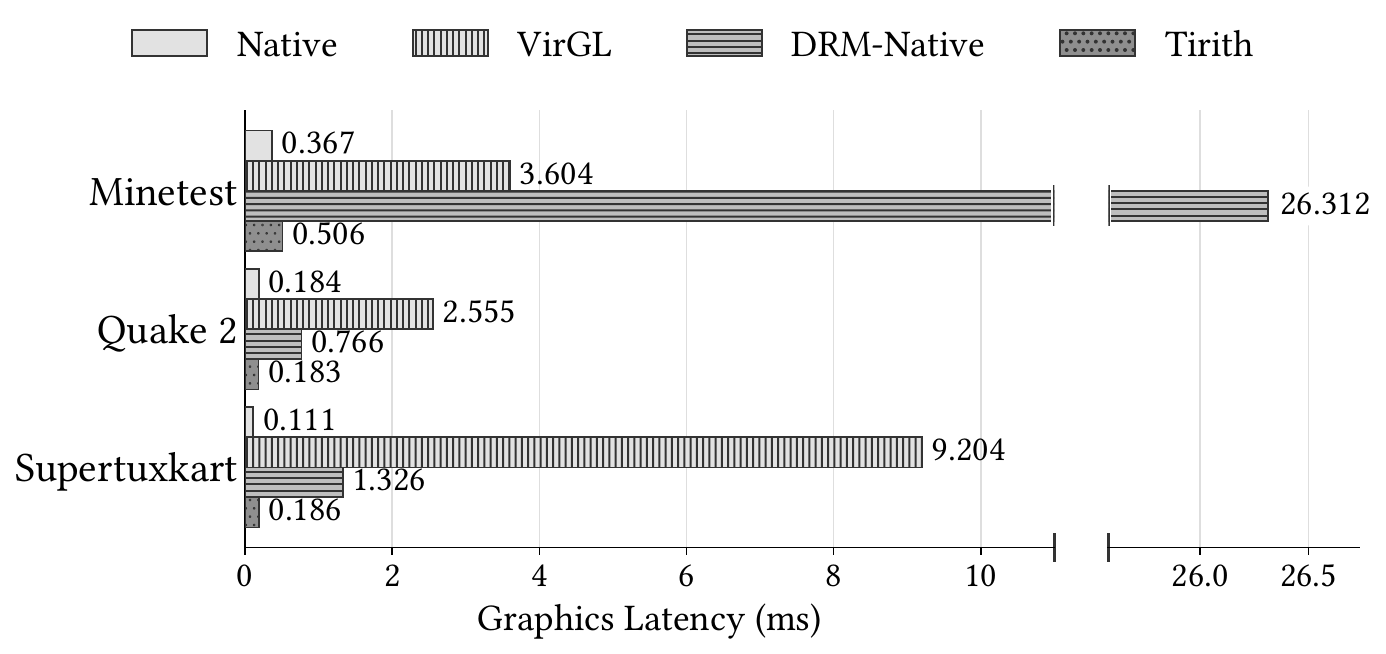}
      \caption {Graphics pipeline latency at 1080p compared across all environments and games.}
     \label{f:graphics-latency-full}
    \end{center}
\end{figure}




%
%

\PP{Initialization and Memory Cost}
\sys incurs a set of initial costs while starting a game session.
\sys must set-up address space shadowing and exception-less message-passing in QEMU ($2.9s$), initialize the LibOS kernel and paravirtualized devices ($56ms$), as well as create rendering contexts~(i.e., set up framebuffers, map GEM objects) and initial window manager contexts ($16ms$)~(\autoref{s:design:rendering}).
%
%
%
Across these initilization steps, \sys only took an average of $3.1s$.
The major cost incurred was for address space shadowing while setting up QEMU, since it requires memory remapping.
%
%
These costs are only one-time and amortized over long-running game sessions.
%
%
In addition, \sys adds minor additional memory overhead for various data structures maintained in the LibOS. 
In our measurements, we found an average of $4.5$-$12.6$MiB memory overhead across games, which is negligible.


\PP{Exception-Less Message Passing Cost}
At runtime, \sys's graphics engine depends heavily on the exception-less message passing rings, therefore we measured the round-trip cost for passing an empty message.
This was measured across $100M$ messages and only took an average of $254{ns}$ per-message. 
This lightweight interface is primarily because message-passing occurs over shared memory and leverages background threads that are woken-up.
%
%

%
%

%

\PP{Additional Input latency}
Input latency is critical for competitive games that have high frame-rates.
\Revision{(R3)}{
Precisely measuring the entire end-to-end latency (e.g., from a physical mouse click to its visible effect on screen) requires external instrumentation such as a high-speed camera, which we currently lack.
However, we evaluated input latency in \sys using virtual mouse clicks for both a game and a custom benchmark.
The clicks are generated by a virtual mouse created through Linux \cc{uinput} and traverse the normal compositor and \cc{X11} input path.

First, we ran Quake 2 and measured the delay from a virtual mouse click into its window until the game processes the input by triggering an attack.
Across 1,000 clicks, \sys{} increases average end-to-end input latency from $584\mu\mathrm{s}$ to $609\mu\mathrm{s}$, resulting in an overhead of $25\mu\mathrm{s}$. 
Of this, $20\mu\mathrm{s}$ comes from delivering the input event to QEMU, encoding it, and forwarding it to Gramine through shared memory.
The remaining $5\mu\mathrm{s}$ comes from delivering the event to Quake 2.
Once delivered to the game, processing the event introduces little to no measurable overhead.

Second, we evaluated the input path with a complementary \cc{GLX} application that continuously polls for input.
Across 1,000 clicks, \sys{} increases average end-to-end input latency from $105\mu\mathrm{s}$ to $178\mu\mathrm{s}$, resulting in an overhead of $73\mu\mathrm{s}$. 
Nearly all of this overhead comes from delivering the input event to QEMU, encoding it, and forwarding it to Gramine LibOS through shared memory, while subsequent delivery to the application and event processing introduce little to no measurable overhead.
By continuously polling, the experiment exposes virtualized input latency that could otherwise be masked by a game's less frequent input polling.
%
%

%
%

}

\begin{figure*}[t]
  \centering
  \footnotesize
  \includegraphics[width=\textwidth]{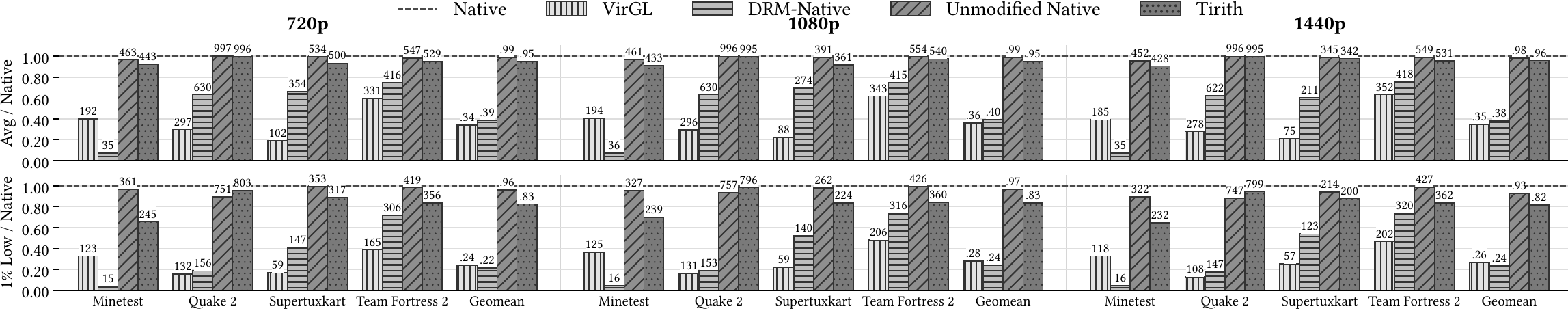}
  \caption{Performance relative to Native 
  \Revision{(R5)}{and Unmodified Native}
  %
  shown in terms of average FPS and 1\% lows. 
  1\% lows reflect frame-time consistency and the severity of FPS dips. Geomean bar labels show relative values; other bar labels show raw FPS.}
  \label{f:evaluation}
\end{figure*}

\subsection{Real-World Video Game Performance}
\label{s:perf:games}

\PP{Game Selection and Benchmarks}
In terms of game selection, we opted to evaluate all environments on a suite of open-source games with OpenGL rendering back-ends and one closed-source game. 
This allows for a deeper understanding of the current performance limitations of \sys{} and existing solutions in well known and highly auditable systems.

%
We selected four games to evaluate the performance of \sys{} against existing paravirtualization solutions.
First, we chose Minetest, a voxel-based, randomly generated open world game, a style of game notorious for stressing systems on high settings.
Second, we chose Quake 2, a classic first-person shooter which demands low latency feedback for the player to make fast-paced combat decisions.
Third, we selected Supertuxkart which is a single or multi-player kart racing game.
\Revision{(R10)}{Finally, we selected Team Fortress 2 which is a competitive online first-person shooter that leverages a modern anti-cheat system called VAC~\cite{steam_vac} which is used in popular competitive games like Dota 2 and Counter-Strike 2. 
We disabled VAC to avoid the system being flagged.
}

For the purposes of selecting benchmarks for our selection of games, Quake 2 and Supertuxkart offer built-in benchmarking features in which either the computer plays the game (Supertuxkart), or the game plays back recorded gameplay sequences (Quake 2). For Minetest, a constant seed and spawn point were chosen for the world, in which the player was spawned in and sat still. 
\Revision{(R10)}{For Team Fortress 2, the player was spawned in a consistent spot on a single map and sat still while the environment changed}.

%

The games were ran on at or near maximum graphics settings for at least 5 minutes, with the data presented being the average of all data at the 5 minute mark. This amount of time allows the games to finish loading assets and in the case of minetest, finish generating local world data. Such concurrent loading operations run alongside rendering, and allowing time for their completion and maximizing graphics settings emphasizes rendering performance.


\PP{Results and Breakdown Analysis}
%
In terms of gaming performance, as shown in \autoref{f:evaluation} \sys{} outperformed all existing paravirtualization solutions, achieving only $4.1$-$5.2\%$ geometric mean overhead compared to \textbf{Native} in average FPS
which is $2.4$-$2.8\times$ better than other solutions. This positive trend continues in $1\%$ lows, in which \sys{} sees $16.5$-$18.2\%$ overhead compared to \textbf{Native}, which is $3$-$3.8\times$ better compared to existing work.
%

%
Starting with Supertuxkart and Quake 2, we see that \sys{} sees $1.3$-$7.4\times$ better average FPS and $1\%$ lows than prior works, incurring only $0.1$-$8.5\%$ overhead compared to \textbf{Native}. 
 Even in the worst case across these games, a user may experience only a minor loss of $11.1$-$16.3\%$ overhead in $1\%$ lows when compared to \textbf{Native}. 
\Revision{(R10)}{\sys{} showcases similar performance and viability when utilized to run the competitive and modern shooter Team Fortress 2.
With average FPS and $1\%$ lows that incur only $2.7$-$4.9\%$ and $15.8$-$16.5\%$ overhead compared to \textbf{Native} respectively. 
This results in achieving a  $1.1$-$2.2\times$ better performance than existing work.}


%
Particularly noteworthy are the Minetest results. \sys{} makes a solid showing with average FPS and $1\%$ lows, only $7.7-9.4\%$ and $30.1$-$35.4\%$ overhead compared to \textbf{Native} respectively.
 This results in achieving a staggering result of $1.9$-$16.4\times$ better performance than prior solutions.
Another observation is that DRM-Native performs very poorly ($4$-$7.5\%$ of \textbf{Native}), when compared with its prior results, it is a clear outlier that requires further investigation.

\Revision{(R5)}{ Analyzing \textbf{Native} versus \textbf{Unmodified Native}, we see worthwhile gains of $1.1$-$1.8\%$ and $3.5$-$8.1\%$ in average FPS and 1\% lows respectively.
%
%
This increased performance is attributed to the EGL backend changes described in~\autoref{s:impl}.
As a result, we see \sys{} close the gap even further to \textbf{Unmodified Native}---bringing the performance overhead to only $2.4$-$4.1\%$ and $11.5$-$13.8$ for average FPS and 1\% lows, respectively, for our tested games.
}


Additional analysis was done on \textit{Graphics Latency}~(\autoref{f:graphics-latency-full}). We define \textit{Graphics Latency} as graphics pipeline overhead including CPU overhead from copying data and sending encoded data across a paravirtualized interface~(\autoref{s:study:preliminary}). \sys{} only sees a maximum increase of $0.14ms$ over \textbf{Native}, other solutions lag behind substantially at a maximum latency increase of $25ms$. The excellent performance of \sys{} is due to the zero-copy nature of the Native Shared GEM Context~(\autoref{s:design:rendering}), which minimizes CPU overhead in the graphics pipeline incurred by existing solutions.

%
%

\AD{Take a pass}

\subsection{Key Takeaways}
%


%
%

\begin{packeditemize}
    \item \sys{} achieves average FPS and $1\%$ lows within $4.1$-$5.2\%$ and $16.5$-$18.2\%$ of \textbf{Native} and $2.4$-$4.1\%$ and $11.5$-$13.8\%$ of \textbf{Unmodified Native}.
    
    \item \sys{} performs $2.4$-$2.8\times$ and $3$-$3.8\times$ better than existing paravirtualization solutions in average FPS and $1\%$ lows. 
\end{packeditemize}

%

Note that existing anti-cheat solutions already incur performance overhead within the range of 5-10\% and even large amounts of system instability as reported by users across many games \cite{blizzard_forums_anticheat_2026, vanguard_performance, faceit_ac_fps_drop, rust_eac_bluescreen}. 
In that regards, \sys's overhead is similar to kernel-level anti-cheats, while offering strong user privacy as outlined in \autoref{s:security:privacy}. 
Moreover, \sys provides equivalent protections against cheating mechanisms as compared to kernel-level solutions.
Therefore, \sys{} as an end-to-end anti-cheat solution, is a compelling approach to address the gap between user privacy and cheat prevention.

%

%% file: relwk.tex

%

\section{Related Work}
\label{sec:related}

\AD{TODO: Adil}

\sys is part of a broader line of work on isolating code that runs on a machine whose root user is untrusted, without granting that code unrestricted host privilege.
This problem has been approached over the years with progressively coarser isolation primitives, and we organize the relevant prior work along that trend.
\SL{better to focus on two-way sandboxes?} \CQ{fix}

The first trend used user-space {\em trusted execution environments} (TEEs), most notably Intel SGX.
Haven~\cite{haven} demonstrated how to shield unmodified Windows applications inside an enclave.
Graphene-SGX~\cite{graphene-sgx}, Panoply~\cite{panoply}, and Scone~\cite{scone} extended the same approach to Linux applications and containers.
%
%
Ryoan~\cite{ryoan} built a distributed sandbox for untrusted computation on secret data.
These designs share a common shape: the protected workload runs in a user-space enclave that is isolated from a compromised host kernel.
However, they cannot themselves provide kernel-level functionality such as driver admission or anti-debugging, and so cannot replace a privileged kernel anti-cheat on the host.

A more recent trend on consumer machines moves the isolation boundary up to whole-VM virtualization.
Microsoft Virtualization-Based Security (VBS) Enclaves~\cite{vbs-enclave} use the Hyper-V hypervisor to isolate sensitive workloads from a host Windows kernel, and Android Protected KVM~\cite{pkvm,pkvm-android} provisions Protected VMs on consumer phones with hardware-enforced two-way isolation.
%
TwinVisor~\cite{twinvisor} achieves the same property on ARM by leveraging TrustZone.
%
%


%% file: conclusion.tex
\section{Conclusion}
\label{s:conclusion}
We presented \sys, a virtualization-based architecture for private and secure PC gaming that replaces today's kernel-level anti-cheats with a two-way Protected VM.
By splitting responsibility between the host's first-party virtualization monitor and the game developer's code, \sys preserves the full set of enforcement primitives kernel anti-cheats provide today while removing the host-side privacy concern.
With the native shared-GEM rendering and bridged split-windowing pipelines, \sys delivers near-native graphics and display performance.
Our evaluations on real games and cheating mechanisms show that \sys can achieve strong security guarantees with modest performance overheads.

%

\AD{Discuss looking glass somewhere here. This approach takes two GPUs (e.g., Intel and Nvidia). It passes the slower GPU to the VM and captures the frame at the driver-level (inside the VM). Then, it leverages a frame transfer operation to copy data to the faster GPU (on the host). This approach has similarities to our idea, but requires multiple dGPUs. It is likely also a bit slower as far as I can tell.}

\SL{some comparisons against cloud gaming (stadia, xbox cloud) in terms of security (in terms of performance would be so trivial) might be interesting. cloud gaming does all rendering in the cloud and transmits video streams to each client. the security of GL commands might be comparable to video stream (if both GL renderer and video stream decoder have no serious bugs). In terms of input protection, cloud gaming might still have to rely on privileged monitoring software. }

\section*{Acknowledgements}
We thank our shepherd and reviewers for their insightful comments.
This work was partly supported by the U.S. National Science Foundation (NSF) under Cooperative Agreement No. 2543639.\AD{any more?}